\documentclass[5p]{elsarticle}
\usepackage{graphicx}
\usepackage{amsmath,amsfonts,amssymb}
\usepackage{bm}
\usepackage{algorithm}
\usepackage{algorithmic}
\usepackage{url}
\usepackage[breaklinks=true,letterpaper=true,colorlinks,bookmarks=false]{hyperref}
\usepackage[table]{xcolor}

\usepackage[pagewise]{lineno}

\biboptions{sort}

\definecolor{deep_blue}{rgb}{0,.2,.5}
\definecolor{dark_blue}{rgb}{0,.15,.5}

\hypersetup{pdftex,  
  breaklinks=true,  
  colorlinks=true,
  linkcolor=dark_blue,
  citecolor=deep_blue,
}

\def\slantfrac#1#2{\kern.1em^{#1}\kern-.1em/\kern-.1em_{#2}}

\newif\iffinal
\finaltrue
\iffinal\else
\usepackage[tightpage,active]{preview}
\fi

\usepackage{xcolor}
\usepackage[normalem]{ulem}
\colorlet{markercolor}{purple!50!black}

\usepackage{MnSymbol}
\begin{document}

\title{Learning and comparing functional connectomes across subjects}

\author[parietal,unicog,cea]{Ga\"el Varoquaux\corref{corresponding}}
\author[cmi,nki]{R. Cameron Craddock}

\cortext[corresponding]{Corresponding author}

\address[parietal]{Parietal project-team, INRIA Saclay-\^ile de France}
\address[unicog]{INSERM, U992}
\address[cea]{CEA/Neurospin b\^at 145, 91191 Gif-Sur-Yvette}
\address[cmi]{Child Mind Institute, New York, New York}
\address[nki]{Nathan Kline Institute for Psychiatric Research, Orangeburg, New York}

\begin{abstract}
	Functional connectomes capture brain interactions via synchronized
	fluctuations in the functional magnetic resonance imaging signal. If
	measured during rest, they map the intrinsic functional architecture of
	the brain. With task-driven experiments they represent integration
	mechanisms between specialized brain areas. Analyzing their variability
	across subjects and conditions can reveal
	markers of brain pathologies and mechanisms underlying cognition.
	Methods of estimating functional connectomes from the imaging signal
	have undergone rapid developments and the literature is full of diverse
	strategies for comparing them. This review aims to clarify links across
	functional-connectivity methods as well as to expose different steps 
	to perform a group study of functional connectomes.
\end{abstract}

\begin{keyword}
    Functional connectivity, connectome, group study, effective
    connectivity, fMRI, resting-state
\end{keyword}

\maketitle

\sloppy 

\section{Introduction}

Functional connectivity reveals the synchronization of distant neural systems
via correlations in neurophysiological measures of brain activity 
\cite{friston1993, biswal1995}.  Given that
high-level function emerges from the interaction of specialized units
\cite{tononi1992},
functional connectivity is an essential part of the description of brain
function, that complements the localizationist picture emerging from the
systematic mapping of regions recruited in tasks \cite{sporns2004}. However,
while there exists a well-defined standard analysis framework for activation
mapping that enables statistically-controlled comparisons across subjects
\cite{friston1995}, group-level analysis of functional connectivity still face
many open methodological challenges. Deriving a picture of a single subject's
functional connectivity is by itself not straightforward, as the brain comprises
a myriad of interacting subsystems and its connectivity must be decomposed into
simplified and synthetic representations. An important view of brain
connectivity is that of distributed functional networks depicted by their
spatial maps \cite{fox2005}. Another no less important and complementary view is
that of connections linking localized functional modules depicted as a graph
\cite{bullmore2009}. This representation of brain connectivity is often called
the functional connectome \cite{sporns2005} and is the focus of intense
worldwide research efforts as it holds promises of new insights in cognition and
pathologies \cite{greicius2008b,biswal2010,fox2010}.

The purpose of this paper is to review methodological progress in the
estimation of functional connectomes from blood oxygenation level dependent
(BOLD) based functional magnetic resonance imaging (fMRI) data and their
comparisons across individuals.  It does not attempt
to be exhaustive, as the field is wide and moving rapidly, but details
specific tools and guidelines that, in the experience of the authors,
lead to controlled and powerful inter-subject comparisons. The paper is
focused on functional connectomes in contrast to structural connectomes,
as the inference of functional connectivity requires important statistical
modeling considerations that are vastly different from the complications involved
with estimating structural connectivity.
While the notion of functional
connectomics is often associated with the study of resting state
\cite{biswal2010}, the methods presented in this paper are also relevant
for task-based studies. On the other hand, the paper has a focus on fMRI;
although the core concepts presented can be applied to 
magnetoencephalography (MEG) or electroencephalography (EEG)
\cite{stam2004}, additional specific problems such as source
reconstruction must be considered \cite{schoffelen2009}.

``Functional connectivity'' is defined as a measure of synchronization in 
brain signals \cite{friston1994}. More generally, it is interesting as a 
window on underlying synchrony on neural processes \cite{lee2003}.
By ``functional connectome'', here we specifically denote a graph representing
functional interactions in the brain, where the term ``graph'' is taken in its
mathematical sense: a set of \emph{nodes} connected together by \emph{edges}.
Graph nodes (brain regions) correspond to spatially-contiguous and 
functionally-coherent patches of gray matter and edges describe long-range synchronizations between nodes
that are putatively subtended by large fiber pathways \cite{marrelec2006b}.  A
graph can be weighted or not, and is completely equivalent to its
\emph{adjacency matrix}, a symmetric matrix tabulating the connection weights
between each pair of nodes.  Functional-connectivity graphs are used to
represent evoked activity, as in task-response studies \cite{mcintosh2000}, as
well as ongoing activity, present in the absence of specific tasks or in the
background during task and often studied in so-called \emph{resting state}
experiments \cite{raichle2010}. Another important notion that arises from the
study of distributed modes of brain function is that of specialized functional
networks\footnote{In neuroimaging, the term network is sometimes used to denote
	a graph of brain function. To disambiguate the notion of segregated
	spatial mode \cite{fox2005} from that of connectivity graphs, we will
	purposely restrict its usage in this paper.} \cite{fox2005}. With our
definition of the functional connectome, functional networks are not directly
building blocks of the connectome but appear as a consequence of the graphical
structure \cite{varoquaux2010c,varoquaux2012}.

The paper is organized as follows. First we discuss estimation of functional
connectomes. This part, akin to a first-level analysis in standard activation
mapping methodology, is not in itself a group-level operation, but it is a
critical step for inter-subject comparison.  In a following section, we discuss
several strategies for comparing connectomes across subjects. Finally we discuss
the links between the representation of brain connectivity as graphs of
functional connectivity and more complex models, such as effective-connectivity
models.


\section{Estimating functional connectomes}

Here we discuss the inference of connectomes from functional brain
imaging data. We start with preprocessing considerations, followed by
the choice of nodes \emph{i.e.}\ regions,
signal extraction, and the estimation of graphs.  

\subsection{Preprocessing considerations}

In addition to standard preprocessing performed for task-based analysis
(slice-timing correction, realignment, spatial normalization, and possibly
smoothing), connectivity-based analysis require additional denoising 
to separate intrinsic activity from confounding signals. This process involves
regressing time series capturing sources of structured noise from the fMRI data.
Physiological noise due to cardiac and respiration are two important
noise signals \cite{hu1995, lund2006, birn2006, birn2008} that are
difficult to control for and as a result are not commonly regressed out. Instead
the mean signal from white matter (WM) and cerebrospinal fluid (CSF) are used as
surrogates to measure these sources of noise as well as other scanner induced
signal fluctuations \cite{fox2005, lund2006}. More complex models account for
spatial variation in noise by incorporating voxel-specific regressors of
neighboring WM (ANATICOR \cite{jo2010}) or the top components from a
principal
components analysis of high-variance signals (CompCor \cite{behzadi2007}). Head motion
induced signal fluctuations are accounted for by incorporating movement
parameters \cite{friston1996, fox2005, lund2006}. The global mean time series
has been proposed as an additional noise regressor that appears to improve the
spatial specificity of connectivity results \cite{fox2005, fox2009}. This practice
has become controversial since the global signal regression introduces negative correlations
\cite{murphy2009,chang2009,saad2012}. 
Removing these sources of
nuisance in addition to linear trends results in more contrasted
correlation matrices that improve the delineation of functional structures
(fig.\,\ref{fig:correlation_matrices}). 

Filtering to remove high frequencies is often performed, based on the initial
observation that fluctuations implicated in resting-state functional
connectivity are predominately slower than 0.1\,Hz \cite{cordes2001,biswal1995}.
While high-pass and low-pass filtering decrease the impact of some confounds,
recent studies have shown that connectivity is present across the
full spectrum of observed frequencies \cite{smith2012,vanoort2012}. Regressing
out a good choice of confound signals is more specific than frequency
filtering,
and in our experience gives more contrasted correlation matrices\footnote{Note
	that naive use of filtering can induce spurious correlations
	\cite{davey2012}.}.
In addition, the recent developments of very rapid acquisition 
protocols prevent aliasing of the physiological noise with the neural 
signal and give access to more specific noise confounds than traditional
low-TR sequences \cite{boyaciouglu2012}.

It is important to keep in mind that the proposed correction strategies
are approximate and not definitive techniques. This has become particularly
apparent for head motion with reports that micromovements on the scale
of $\le 0.2$\,mm can induce artefactual group-level findings even when motion is
accounted for in preprocessing
\cite{vandijk2012,power2011,satterthwaite2013}. Special care must be taken to
adequately control for residual impact of head motion in the group model
\cite{satterthwaite2013,vandijk2012}.

\subsection{Defining regions}

The choice of regions of interests (ROIs) that define the nodes of the
graphs can be very important both in the estimation of connectomes and
for group comparison \cite{wang2009}. Unsurprisingly, simulations have
shown that extracting signal from ROIs that did not match functional
units would lead to erronous graph estimation \cite{smith2011}.
Different strategies to define suitable ROIs coexist. While dense 
parcellation approaches cover
a large fraction of the brain \cite{achard2006, varoquaux2010c,
wang2009,bellec2006,craddock2012}, this coverage can be traded off to focus on some specific
regions, in favor of increased functional specificity and thus better
differentiation across networks \cite{greicius2003, dosenbach2006,
varoquaux2010b}. In addition, while
ROIs are most often defined as a hard selection of voxels, it is also
possible to use a \emph{soft} definition, attributing weights as with
probabilistic atlases, or spatial maps of functional networks extracted
from techniques such as independent component analysis (ICA) \cite{kiviniemi2009,smith2012}.

\begin{figure}
\iffinal
    \includegraphics[width=\linewidth]{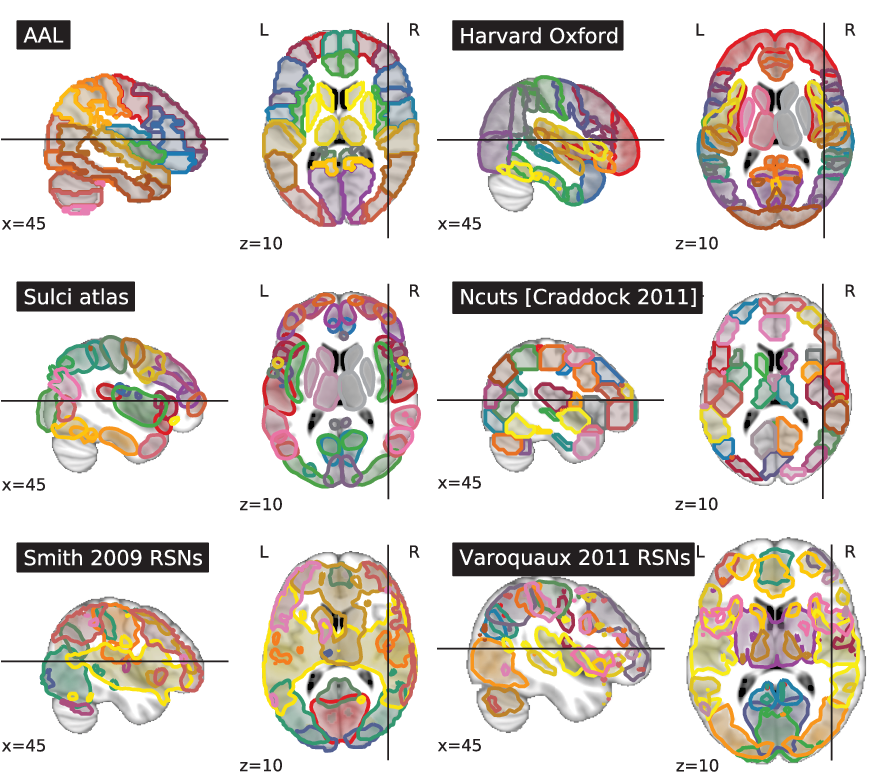}%
\fi

\caption{
Different full-brain parcellations: the AAL atlas
\cite{tzourio-mazoyer2002a}, the Harvard-Oxford atlas, the sulci atlas used in
\cite{varoquaux2010c}, regions extracted by Ncuts
\cite{craddock2012}, the resting-state networks extracted in
\cite{smith2009} by ICA, and in \cite{varoquaux2011} by sparse dictionary
learning.
\label{fig:parcellations}
}
\end{figure}

\paragraph{Regions from atlases}
Atlases can be used to define full-brain parcellations. Popular
choices are the Automatic Anatomic Labeling (AAL) atlas \cite{tzourio-mazoyer2002a}, 
which benefits from an SPM toolbox, or the ubiquitous Talaraich-Tournoux
atlas \cite{talairach1988}. 
However, these atlases suffer from major
shortcomings; namely \emph{i)} they were defined on a single subject
and thus do not reflect inter-subject variability, and \emph{ii)}
they focus on labeling large anatomical structures and do not match
functional layout --for instance only two regions describe the medial
part of the frontal lobe in the AAL atlas. Multi-subject probabilistic altases such as the
Harvard-Oxford atlas distributed with FSL \cite{smith2004} or the
sulci-based structural atlas used in \cite{varoquaux2010c} mitigate the
first problem, and the high number of regions defined using sulci also
somewhat circumvent the second problem (see fig.\,\ref{fig:parcellations}).

\paragraph{Defining regions from the literature}
Regions can be defined from previous studies, informally or with
systematic meta-analysis. This strategy is used to define the main
resting-state networks, such as the default mode network, but may also be
useful to study connectivity in task-specific networks
\cite{biswal1995,rissman2004,dosenbach2006,grillon2012}. The common practice is to place balls of a given
radius, 5 or 10\,mm, centered at the coordinates of interest. Given that
functional networks are tightly interleaved in some parts of the cortex,
such as the parietal lobe, care must be taken not to define too many
regions that would overlap and lead to mixing of the signal.

\paragraph{FMRI-based function definition}
Defining regions directly from the fMRI signal brings many benefits.
First, it can capture subject-specific functional information. Second, it
adapts to the signal at hand and its limitations, such as
image distortions or vascular and movement artifacts that are isolated
in ICA-like approaches. Lastly, incorporating functional information into
regional definition will result in more homogenous regions that better represent
connectivity present at the voxel level than anatomically-defined atlases such
as AAL or Harvard-Oxford \cite{craddock2012}. The simplest approach to define task-specific
regions is to use activation maps derived from standard GLM-based
analysis in a task-driven study (see for instance \cite{poldrack2011}).
Regions are extracted by thresholding the maps, or using balls around the
activation peaks. For resting-state studies, unsupervised multivariate
analysis techniques are necessary. Clustering approaches extract
full-brain parcellations \cite{craddock2012,bellec2010,yeo2011,thirion2006}, and have been shown to segment well-known functional
structures from rest data. Alternatively, decomposition
methods, such as ICA \cite{beckmann2004}, can unmix linear
combinations of multiple effects and separate out partially-overlapping
spatial maps
that capture functional networks or confounding effects, as for instance
with the presence of vascular structure in functional networks. At high model
order, ICA maps define a functional parcellation \cite{kiviniemi2009}.
Extracting regions from these maps requires additional effort as they can
display fragmented spatial features and structured background noise, but
incorporating sparsity and spatial constraints in the decomposition
techniques leads to contrasted maps that outline many different
structures \cite{varoquaux2012} (see fig.\,\ref{fig:parcellations}).

\paragraph{Optimal number of regions}
Defining an optimal number of regions to use for whole-brain connectivity analysis bears
careful consideration. On one hand we desire a sufficiently large number of
regions to guarantee that they are functionally homogeneous regions and adequately 
represent the connectivity information present in the data. On the other hand
too many regions will render statistical inference challenging, 
result in an explosion in computational complexity,
and interfere with the interpretability of observed connections. For functional
parcellation, cross-validation methods can be employed to estimate an optimal 
number of regions based on homogeneity, the ability to reproduce 
connectivity information present at the voxel scale, and the ability to
obtain the same parcellations from independent data 
\cite{craddock2012, blumensath2012}. In general these metrics do not result in an
obvious peak at a ``best'' number of regions, but instead offer a range over
which the number of regions can be chosen based on the needs of the analysis at
hand. Finally, it is important to keep in mind that there is no
universally better parcellation and associated number of regions. From a
practical standpoint, these choices will depend on the task at hand, and
more fundamentally, a good description of brain function should cover
multiple scales. Given that it is not clear that an optimal parcellation
can be identified from the sample size of a typical study, randomized
parcellation, as used in structural connectomes \cite{zalesky2010a} or
activation mapping \cite{varoquaux2012a}, may also be considered.

\subsection{Estimating connections}

The concept of functional connectivity has been called elusive
\cite{horwitz2003}: it has many mathematical instantiations although in
essence they all strive to extract simple statistics from functional
imaging in order to characterize synchrony and communication between large
ensembles of neurons. Here we choose to focus on second order statistics
that can be related to Gaussian models, the simplest of which being the
correlation matrix of the signals of the different ROIs.

\begin{figure}
\iffinal
    \includegraphics[width=\linewidth]{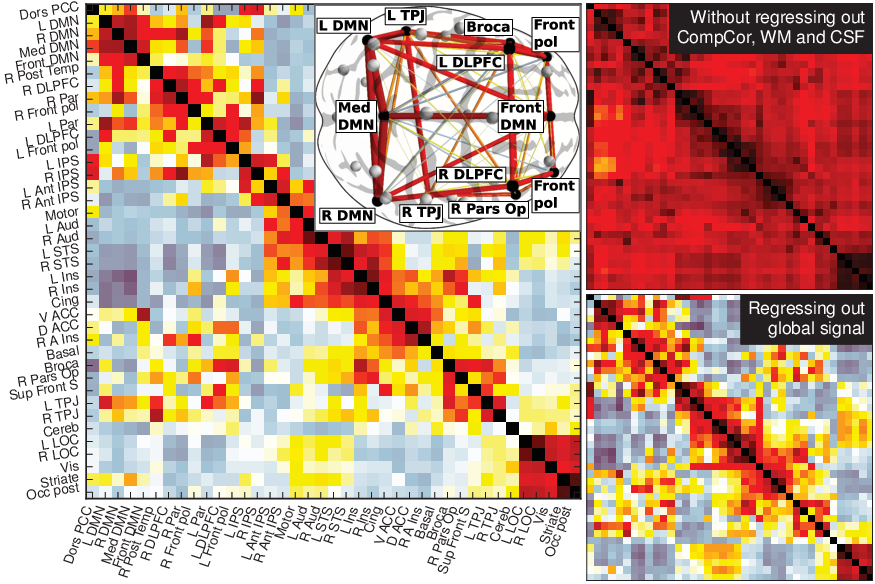}%
\fi

\caption{
Correlation matrices of rest time-series extracted from the 39 main
regions of the Varoquaux 2011 \cite{varoquaux2011} parcellation with
different choices of confound regressors -- 
\textbf{Left}: regressing out CompCor signals, as well as white matter and
CSF average signals and movement parameters. The insert shows the
connections restricted to a few major nodes.
-- \textbf{Upper right}: regressing out only movement parameters. -- 
\textbf{Lower right}:
regressing out movement parameters and global signal mean.
No frequency filtering was applied here.
\label{fig:correlation_matrices}
When no confounding brain signals are regressed, all regions are heavily
correlated. Regressing out common signal, in the form of well-identified
confounds or a global mean, teases out the structure.
}
\end{figure}

\paragraph{Signal extraction}
Given a set of graph nodes, the next step is to extract a representative time
series for each node.  To study \emph{intrinsic} activity, \emph{e.g.}\ with
rest data, signal extraction can be achieved by either averaging the fMRI time
series across the voxels in a region, or by taking the first eigenvariate from a
principle components analysis of the time series \cite{friston2006}.
Comparisons of these methods has shown that the eigenvariate method is
more sensitive to function inhomogeneity \cite{craddock2012} and exhibits worse
test-retest reliability than averaging time series \cite{zuo2010}. 
In addition, improved specificity to BOLD signal can be enforced
by using only signal in voxels near gray-matter tissues. For this purpose, we
suggest summarizing the signal in an ROI by a mean of the different voxels
weighted by the subject-specific gray matter probabilistic segmentation, 
as output by \emph{e.g.}\ SPM's segmentation tool \cite{ashburner2005} or
FSL's FAST program \cite{zhang2001}.

Studying connectivity from \emph{evoked} activity with task-driven studies 
requires disambiguating task-specific connectivity effects from
intrinsic connectivity mediated by shared neuromodulatory/task 
inputs, anatomical pathways, \emph{etc}. In this regard, it
can be beneficial to run a GLM-based first-level analysis, enforcing specificity
of the measure extracted to the task. With slow event-related designs,
task-specific functional connectivity can be captured in trial-to-trial
fluctuations in the BOLD response, estimated using a GLM analysis with one
regressor per trial \cite{grillon2012,rissman2004,mennes2010}. This approach,
known as beta-series regression, has been adapted for rapid event-related
designs, using multiple GLMs to optimize deconvolution of each trial
\cite{mumford2012}.

\begin{figure}
\iffinal
    \includegraphics[width=\linewidth]{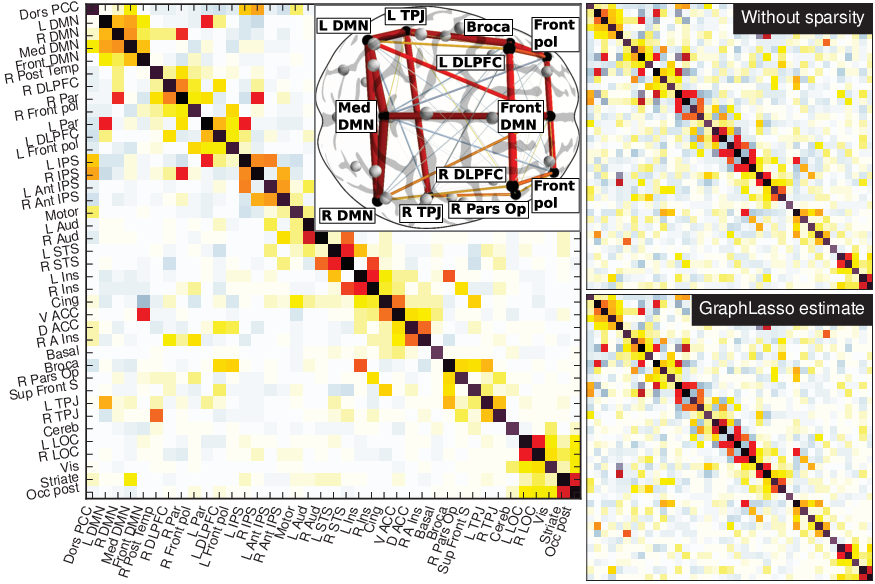}%
\fi

\caption{
Different inverse-covariance matrices estimates corresponding to
fig.\,\ref{fig:correlation_matrices} -- \textbf{Left}: group-sparse estimate
using the $\ell_{21}$ estimator \cite{varoquaux2010c}.
The insert shows the
connections restricted to a few major nodes. -- \textbf{Upper
right}: non-sparse estimate: inverse of the sample correlation matrix. --
\textbf{Lower right}: sparse estimate using the Graph Lasso
\cite{friedman2008}.
\label{fig:icov_estimators}
}
\end{figure}

\paragraph{Correlation and partial correlations}
Given ROIs defining the nodes of the functional-connectome graph,
one needs to estimate the corresponding edges connecting them.
Functional connectivity between the ROIs can be measured by computing the
correlation matrix of the extracted signals. An important and often
neglected point is that the sample correlation matrix, \emph{i.e.}\ the
correlation matrix obtained by plugging the observed signal in the
correlation matrix formula, is not the population correlation matrix,
\emph{i.e.}\ the correlation matrix of the data-generating process. If the
number of measurements was infinite, the two would coincide, however if
this number is not large compared to the number of connections (that
scales as the square of the number of ROIs), the sample correlation
matrix is a poor estimate of the underlying population correlation
matrix. In other words, the sample correlation matrix captures a lot of
sampling noise, intrinsic randomness that arises in the estimation of
correlations from short time series. Conclusions drawn from the sample
correlation matrix can easily reflect this estimation error.
Varoquaux \emph{et al.}~\cite{varoquaux2010c} and Smith \emph{et
al.}~\cite{smith2011} have shown respectively on rest fMRI and on
realistic simulations that a good choice of correlation matrix estimator
could recover the connectivity structure, where the sample correlation
matrix would fail.  In general, the choice of a better estimate depends on
the settings and the end goals \cite{varoquaux2012,varoquaux2010b},
however the Ledoit-Wolf shrinkage estimate \cite{ledoit2004} is a simple,
computationally-efficient, and parameter-less alternative that performs
uniformly better than the sample correlation matrix
\cite{varoquaux2012,varoquaux2010c} and should always be preferred.

\begin{figure}
\iffinal
    \includegraphics[width=\linewidth]{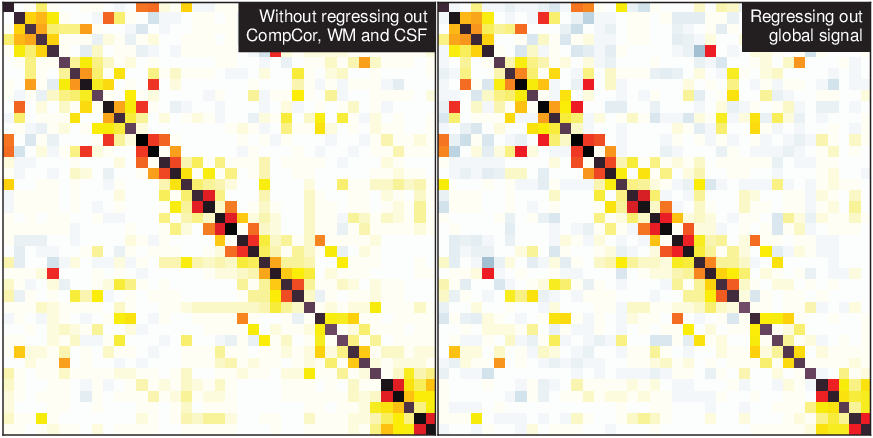}%
\fi

\caption{Inverse-covariance matrices for different choice of
confound regressors --
\label{fig:icov_regressors}
\textbf{Left}: regressing out only movement parameters --
\textbf{Right}: removal of the global mean,
instead of the white matter, CSF, and CompCor time courses. 
}
\end{figure}

For the problem of recovering the functional-connectivity
\emph{structure}, \emph{i.e.}\ finding which region is connected to
which, sparse inverse covariance estimators have been found to be
efficient \cite{varoquaux2010c,smith2011,ryali2012}. The intuition for relying on
inverse covariance rather than correlation stems from that fact that
standard correlation (marginal correlation) between two variables $a$ and
$b$ also capture the effects of other variables: strong correlation of
$a$ and $b$ with a third variable $c$ will induce a correlation between
$a$ and $b$. On the opposite, the inverse covariance\footnote{Covariance
and correlation matrices differ simply by the fact that a covariance
matrix captures the amplitude of a signal, via its variance, while a
correlation matrix is computed on standardized (zero mean, unit variance) signals.}
matrix (also called \emph{precision matrix}) captures
partial correlations, removing the effect of other variables
\cite{marrelec2006a}. In the small sample limit, this removal is
challenging from the statistical standpoint. This is why an assumption 
of sparsity,
\emph{i.e.}\ that only few variables need to be considered at a time, is
important to estimate a good inverse covariance. Various estimation
strategies exist for sparse inverse covariance, and have an impact on the
resulting networks \cite{varoquaux2012,varoquaux2010c}. The
Graph Lasso ($\ell_1$-penalized maximum-likelihood estimator)
\cite{friedman2008} is in general a good approach for structure recovery. In group studies,
the $\ell_{21}$ estimator \cite{varoquaux2010c,honorio2012} is useful to
impose a common sparsity structure across different subjects and achieve
better recovery of this common structure. Simply put, these approaches
are necessary because estimation noise creates a background structure
(see fig.\,\ref{fig:icov_estimators}); however, unlike in a univariate situation, the
parameters are not independent, and the spurious background connections
degrade the estimation of the actual connections. The sparse estimators
make a compromise between imposing simpler models, \emph{i.e.}\ with less
connections, and providing a good fit to the data. This compromise is set via a
regularization parameter which controls the sparsity of the estimate. A
good procedure to choose this parameter is via cross-validation
\cite{varoquaux2010c}.


\paragraph{Network structure extracted}
The correlation matrices and inverse-covariance matrices that we extract
contain a lot of information on the functional structure of the brain.
First, the correlation matrix (fig.\,\ref{fig:correlation_matrices})
shows blocks of synchronized regions that can be interpreted as
large-scale functional networks, such as the default mode network. Note
that the split in networks is not straightforward. Different ordering of the
nodes will reveal different networks. 
Indeed, because of the presence of
hubs and interleaved networks, the picture in terms of segregated networks
is not sufficient to explain full-brain connectivity
\cite{varoquaux2012}. Connectivity matrices, correlation matrices and
inverse-covariance matrices, can be represented as graphs: nodes connected
by weighted edges (inserts on fig.\,\ref{fig:correlation_matrices} and
fig.\,\ref{fig:icov_estimators}). The inverse-covariance matrix, which
captures partial correlations, appears then as extracting a
\emph{backbone} or \emph{core} of the graph.
While such structure has
been used as a way to summarize anatomical brain connectivity graphs
\cite{hagmann2008}, here it has a clear-cut meaning with regard to
the BOLD signal: it gives the conditional independence structure between
regions \cite{varoquaux2012}. In other words, regions $a$ and $b$ are
not connected if the signal that they have in common can be explained by
a third region $c$. In this light, the choice of nuisance regressors to
remove confounding common signal is less critical with partial
correlations than with correlations. Indeed, while with
correlation matrices regressing out the global mean has a drastic effect
(fig.\,\ref{fig:correlation_matrices} upper right and lower right), on
inverse covariance it only changes the resulting matrices very slightly 
(fig.\,\ref{fig:icov_regressors}).

There have been debates on whether to
regress out certain signals, such as the global mean,  as it induces
negative correlations \cite{murphy2009,chang2009,fox2009}, and these may seem
surprising: one network appears as having opposite fluctuations to
another. However,
correlation between two signals only takes its meaning with the
definition of a baseline. A simple picture to explain anti-correlations
between two regions is the presence of a third region, mediating the
interactions. Using this third region as a baseline would amount to
estimate partial correlations in the whole system. Using 
inverse-covariance matrices or partial correlations to understand brain
connectivity makes the interpretation in terms of interactions between
brain regions easier and more robust to the choice of confounds.



\section{Comparing connectivity}

We now turn to the problem of comparing functional
connectivity across subjects or across conditions.

\subsection{Detecting changing connections}

First, we focus on detecting where the connectivity matrices estimated in
the previous section differ.

\paragraph{Mass-univariate approaches}
The most natural approach is to apply a linear model to each coefficient of
the connectivity matrices \cite{lewis2009,grillon2012}. This
approach is similar to the second-level analysis used in mass-univariate
brain mapping, and gives rise to many of the well-known techniques
used in such a context, such as the definition of a second-level design,
with possibly the inclusion of confounding effects, and statistical tests
(T tests or F tests) on contrast vectors. Importantly, in order to work
with Gaussian-distributed variables, it is necessary to apply a Fisher
Z transform\footnote{See
\url{http://en.wikipedia.org/wiki/Fisher_transformation} or
\cite{anderson1958} section 4.2.3 for mathematical arguments.} to the
correlations. Note that
in these settings, the Ledoit-Wolf estimator \cite{ledoit2004} is
often a good choice to estimate the correlation matrix, as it is
parameter-free and gives good estimation performance without imposing any
restrictions on the data.
For hypothesis testing,
correcting for multiple comparison can severely limit statistical
power, as the number of tests performed scales as the square of the
number of regions used. Controlling for the false discovery rate (FDR)
mitigates this problem. Alternatively, as the assumptions underlying the
Benjamini-Hochberg procedure \cite{benjamini1995} for the FDR can easily
be broken, non-parametric permutations-based tests give reliable
approaches. In particular, the max-T procedure \cite{ge2003,nichols2001}
is interesting to avoid the drastic Bonferroni correction when
controlling for multiple comparison in family-wise error rate.

\begin{figure}
\iffinal
    \includegraphics[width=\linewidth]{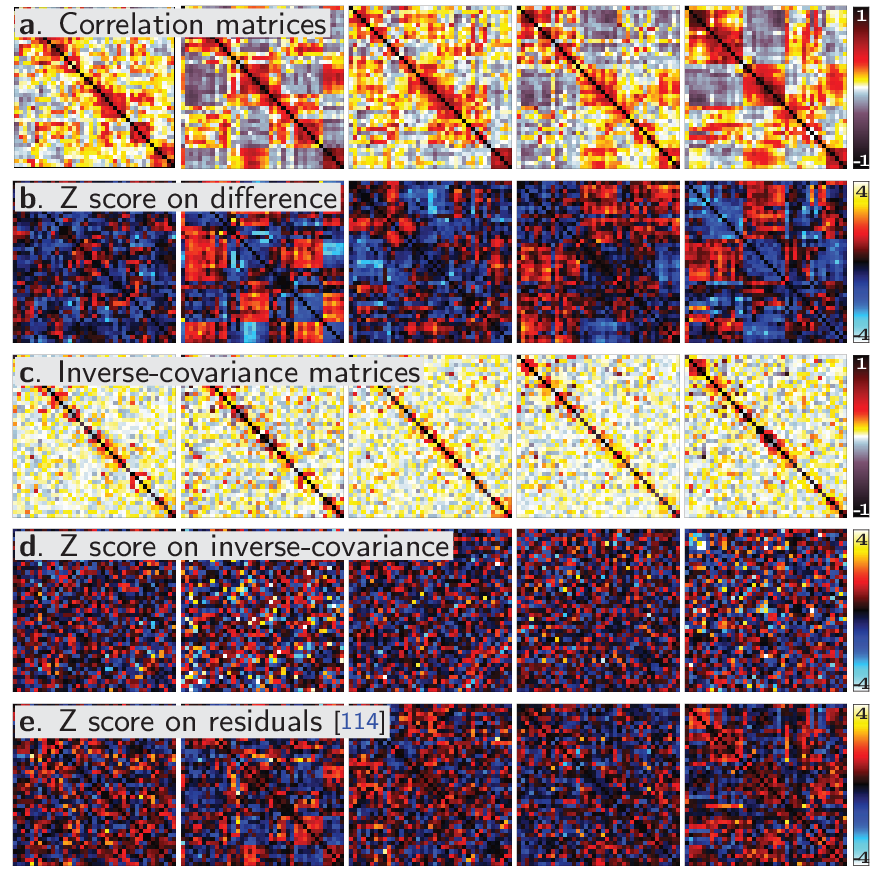}%
\fi

\caption{Inter subject variability. Note that this is variability
occurring in a healthy population at rest, in other words it is non specific
variability -- \textbf{a}: single-subject
correlation matrices for different subjects -- \textbf{b}:
Corresponding Z-score (effect / standard deviation) of the difference
between a subject and the remaining others -- \textbf{c}:
single-subject inverse-covariance matrices -- \textbf{d}: Corresponding Z-score
for the inverse-covariance matrices -- \textbf{e}:
Corresponding Z-score for the subject residuals, as defined in 
\cite{varoquaux2010b}.
\label{fig:inter_subject}}
\end{figure}

\paragraph{Accounting for distributed variability}
A specific challenge of connectivity analysis is that the connectivity
strength between different regions tends to covary. For instance, with
resting-state data, functional networks comprising many nodes can appear
as more or less connected across subjects (see for instance
fig.\,\ref{fig:inter_subject}, showing variability in a control
population at rest). In other words, non-specific variability is
distributed across the connectivity graph, and it is structured by the
graph itself. This observation brings the natural question of whether
second-level analysis should be performed on correlation matrices,
inverse-covariance matrices, or another parametrization that would
disentangle effects and give unstructured (white) residuals. While
inverse-covariance matrices show less distributed fluctuations than
correlation matrices, they capture a lot of background noise, as partial
correlations are intrinsically harder to estimate. Preliminary work
\cite{varoquaux2010b} suggests performing statistical tests on residuals of
a parametrization intermediate between correlation matrix and inverse
covariance matrix, as it can decouple effects and noise.

Taking a different stance on distributed variability, the ``network-based
statistics'' approach \cite{zalesky2010} draws from the hypothesis that
if, in a second-level analysis, an effect is detected on a connection
that lies in a network of strongly connected nodes, a large
sub-network is
likely to carry an effect. Thus, they adapt cluster-level inference to
connectivity analysis, in order to mitigate the curse of multiple
comparisons.


\subsection{Comparing network summary statistics}

Both the multiple comparison issue and the network-level distributed
variability are a plague to edge-level comparison of connectomes. A
possible strategy to circumvent these difficulties is to perform
comparisons and statistical testing at the level of the network, rather
than the individual connection. 

\begin{figure}
\iffinal
    \centerline{\includegraphics[width=.9\linewidth]{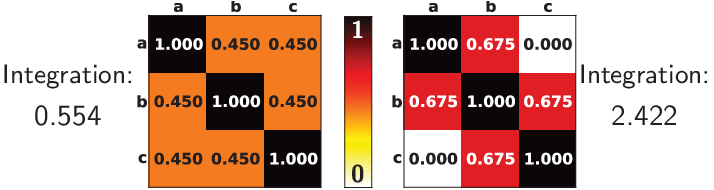}}
\fi

\caption{Two different correlation matrices with the same average
correlation, but with very different integration values. Indeed, the
matrix on the left was chosen to represent three signals $a$, $b$
and $c$ as different from each other as possible, given $\rho_{ab} +
\rho_{bc} + \rho_{ac} = 1.35$; it thus has a small integration value. On
the opposite, for the matrix on the right, signal $b$ can almost be fully
recovered by combining signals $a$ and $c$; the matrix thus has a large
integration value.
\label{fig:integration}}
\end{figure}

\paragraph{Network integration}
Marrelec \emph{et al.}\
\cite{marrelec2008} introduce the use of entropy and mutual information
as a measure of network-level functional integration\footnote{See \cite{varoquaux2010c}
for simplified formulas for network integration and mutual information.}.
Gaussian entropy can be seen as a simple metric to generalize correlation
or variance to multiple nodes (see \cite{anderson1958} \S2.5.2 and
\S7.5). Indeed, let us consider 3 nodes: $a$, $b$ and $c$. Their
correlation structure is captured by three correlation coefficients:
$\rho_{ab}$, $\rho_{bc}$ and $\rho_{ac}$. Summarizing these by their
mean, as might seem natural, discards the relationship between the
signals, while using the integration metric, defined as the Gaussian entropy, tells us how much two signals can be
combined to form the third (see fig.\,\ref{fig:integration}). 
Cross-entropy --or mutual information--
\cite{marrelec2008} measures the amount of cross-talk between two
systems in a similar way as Gaussian entropy is used to measure the
integration of a brain
system. The functional-connectivity structure, or its representation in
the form of a correlation matrix, can thus be characterized via the
integration and cross-talk of some of its sub-systems. This approach
gives a simplified representation with a small number of metrics that can
be compared across subjects.

\paragraph{Graph-topological metrics}
Functional connectivity graphs have been found to display specific
topological properties\footnote{In the neuroscience world, these
descriptions are grouped under the terms of ``graph-theoretical
approaches'', however graph theory is an entire division of mathematics
and computer science that is concerned with much more than topology of
random graphs.} that are characteristic of small-world networks
\cite{stam2004,salvador2005,achard2006,bullmore2009}. These networks
display excellent transport properties: although they have a relatively
small number of connections, any two regions of the brain are well
connected. 
Another interesting consequence of their specific topology is
the resilience it gives the system to attacks such as resulting from
brain lesions \cite{achard2006}. This overall structure of
functional-connectivity graphs can be summarized by a few metrics, such
as the average path length between any two nodes, the local clustering
coefficients, or the node degree centrality \cite{rubinov2010}. Given that pathologies without a
localized focus, such
as schizophrenia, are thought to have a global impact on brain
connectivity \cite{liu2008,bassett2008}, the graph-topological metrics are
promising markers to perform inter-subject comparison. Such an
approach is appealing as it is not subject to multiple comparison issues.
However, it has been criticized as giving a fairly unspecific
characterization of the brain and being fragile to noise
\cite{ioannides2007}. Another caveat is that these properties are not
specific to brain function: correlation matrices display
small-world properties such as local clustering by construction. Indeed, if two
nodes are strongly correlated to a third, they are highly likely to be
correlated to each other \cite{zalesky2012}. This observation highlights
the need for well defined null-hypothesis \cite{zalesky2012,rubinov2011},
but also for controlled recovery of brain functional connectivity going
beyond empirical correlation matrices, as discussed in the previous section.


\subsection{Predictive Modeling}



Predictive modeling is concerned with learning (or fitting) a model that is
capable of \emph{predicting} information from unseen data
\cite{pereira2009}.  In the context of connectomes, predictive modeling
can extract connectivity-based biomarkers of disease diagnosis,
prognosis, or other phenotypic outcomes \cite{craddock2009, dosenbach2010}. The
accuracy of a predictive model provides a measure of the amount of information
present in the connectome about the phenotypic measure being evaluated
\cite{kjems2002, kriegeskorte2006}.  When combined with reproducibility, prediction accuracy
provides a metric for evaluating experimental trade-offs for data acquisition,
preprocessing, and analysis \cite{strother2002, laconte2003}.  Multivariate
predictive models are attractive in connectomics because they are sensitive to dependencies
between features and avoid the need to correct for multiple comparisons since
the significance of an entire pattern is evaluated using a single statistical
test. 
Additionally, modern predictive modeling techniques draw from the
statistical learning literature, which specifically addresses
 high dimensional datasets with few observations.
Predictive modeling has been
successfully applied to identify connectome-based biomarkers of Alzheimer's
disease \cite{stonnington2010}, depression \cite{craddock2009, zeng2012}, schizophrenia
\cite{cecchi2009, shen2010}, autism \cite{anderson2011}, ADHD \cite{zhu2008}, aging
\cite{dosenbach2010}, as well as to classify mental operations
\cite{richiardi2011,shirer2012}. The growing interest for applying
predictive modeling to connectivity analysis was highlighted by the ADHD200
Global Competition, in which the object was to identify a connectivity-based
biomarker of ADHD \cite{adhd2002012}. Recent work has illustrated the utility
of predictive modeling for deriving connectivity models at the individual level
\cite{chu2011}.

Technically, predictive modeling is a supervised machine learning problem where
a target to be predicted --\emph{e.g.}\ age, disease state, cognitive state-- is
available for each observation of the data. In the context of comparing
connectomes, features used in the predictive model correspond to bivariate
measures of connectivity \cite{richiardi2011,dosenbach2010,shirer2012}, or any of the previously discussed graph
summary metrics \cite{cecchi2009,ekman2012}.
The quality of a predictive model is determined by its prediction accuracy (or
generalization ability) which is measured using one or more iterations of
cross-validation. Cross-validation iteratively subdivides available data into a
subset used for training the classifier and a dataset for evaluating classifier
performance\footnote{Several strategies exist for performing cross-validation and the commonly
used approach of using only a single observation for testing (leave-one-out
cross-validation) results in highly variable estimates of prediction accuracy
\cite{friedman2001}.
Alternative approaches such as (5 or 10)-fold cross-validation,
or $0.632+$ bootstrap should be preferred \cite{friedman2001}.} 
 \cite{pereira2009}.
The significance of
achieved prediction accuracy can be assessed using permutation tests
\cite{golland2003}.
Predictive modeling approaches typically require the specification of several
parameters, which may be chosen based on domain specific knowledge or
requirements \cite{cherkassky1998}, determined using an analytical approach
\cite{cherkassky2004}, or optimized using a second-level cross validation procedure
\cite{friedman2001}.  

Although predictive modeling techniques are well suited for measuring whether
information exists in the connectome about a phenotypic variable, they do
not directly identify the connections that are most relevant to the
prediction.  This limitation can be somewhat mitigated  by
relying on previously described sparse inverse covariance estimation techniques to minimize
the number of connections. Additionally, feature selection \cite{guyon2003} 
can be performed by filtering out
features based on their statistical relationship with the variable of interest
\cite{craddock2009, shen2010}. Importantly, it must be performed within
cross-validation to avoid biasing estimates of prediction accuracy.
The interpretation of connections used in predictive models and their
relationship with
a phenotypic outcome is difficult and requires insight into the mechanisms
underlying the modeling approach. For linear models, the weights of the
model are similar to the weights of
an ordinary linear regression. If features are appropriately standardized
prior to training, the magnitude of the weights can be interpreted to reflect
the relative importance of the feature to the model, for instance the
corresponding edge in the connectome. However,
interpreting how the connections differ between classes or relate to a
phenotypic variable can be more complicated given the multivariate nature
of their involvement.
Indeed, the inclusion or exclusion of a connection in the model
can induce a change of the sign of another model weight \cite{craddock2009}.
 It
is perhaps most reasonable to adapt a conservative interpretation in which
predictive modeling is used to identify candidate connections that are later
tested in follow-up experiments better suited to elucidating their relationship
to the variable of interest.

To conclude on predictive modeling with a practical note for connectome
comparison, we would like to stress that while machine learning algorithms are
powerful tools, they work best if they are provided with discriminanting and
noiseless features. In other words, as with all other connectome comparison
methods, optimizing first-level analysis --subject-level connectome extraction--
is paramount.


\section{Beyond correlation, effective connectivity?}

All the approaches that we have presented in this review are based on
second-order statistics of the signal, in other words correlation
analysis. Traditionally, these are defined as \emph{functional
connectivity}, defined as ``temporal correlations between remote
neurophysiological events'' \cite{friston1994}, and opposed to
\emph{effective connectivity}, \emph{i.e.}\ ``the influence one neural
system exerts over another'' \cite{friston1994}. To conclude this review,
we would like to bridge the gap between these concepts, which in our eyes
should be seen as a continuum rather than an opposition (this opinion is
also expressed in \cite{mcintosh2010}).

A first step to move from purely descriptive statistics to interaction
models with
functional connectivity analysis is to consider a correlation matrix as a
Gaussian graphical model, \emph{i.e.}\ a well-defined probabilistic model
that describes observed correlations in terms of an independence
structure and conditional relations \cite{lauritzen1996,varoquaux2012}.
In such settings, the inverse covariance graph or the partial
correlations are a measure of influence from one node to another, albeit
undirected. Inferring directionality in a Gaussian model is impossible. Linear
structural equation models (SEMs) \cite{mcintosh1994} rely on a similar
model that consists in specifying a candidate directed graphical
structure. This structure constraints the covariance matrix of the
signals and can thus be tested on observed data. In fact some forms of
SEMs are known as ``covariance structure models''. There is thus a strong
formal link between correlation analysis in the framework of graphical
models and SEMs: the former is undirected but fully exploratory, as it
does not require the specification of candidate structure, while the
latter is directed but confirmatory. This link has been exploited to
specify candidate structures for SEMs using partial correlations
\cite{marrelec2007}. More complex models, such as dynamical causal models
(DCMs) \cite{friston2003a} or Granger causality \cite{goebel2003} require
additional hypotheses such as non-linear couplings or time lags.

Most importantly, more complex models can only be used to model
interactions between a small number of nodes. This is not only due to a
computational difficulty, but also to fundamental roadblocks in
statistics: the complexity of the model must match the richness of the
data. While injecting prior information can help model estimation, the
more informative this prior is, the more fragile the inference becomes. The
ongoing debate on the impact of hemodynamic lag on Granger-causality
inference \cite{smith2012a} is an example of such fragility.
Note that although most of the theory underpinning correlation analysis
(Gaussian graphical models) is based on a Gaussian assumption, the core results
are robust to violations of this assumption \cite{ravikumar2011}.

It is tempting to favor more neurobiologically-inspired models that give
descriptions close to our knowledge of the brain basic mechanisms,
however, as George Box famously said, ``all models are wrong; some models
are useful''. Depending on the question and the data at hand, a trade-off 
should be chosen between complex models based on a bio-physical description,
and simple phenomenological models such as correlation matrices. In
particular to model interactions between a large number of regions, as in
full-brain analysis, and learn a large \emph{connectome}, simple models
are to be preferred. For more hypothesis-driven studies, such as the
analysis of the mechanisms underlying a specific task, more complex models
can be preferred, if rich data is available. Automatic choice of model is
a difficult problem, however, cross-validation (as used in
\cite{varoquaux2010c,craddock2012,strother2006}) is a useful tool. The
central principle of cross-validation is to test a model on different
data than the data used to fit the model. Models too complex
for the data available will fit noise in the data, and thus generalize
poorly. The main benefit of cross-validation is that it is a
non-parametric method which does not rely strongly on modeling
assumptions\footnote{This is to be contrasted to Bayesian model
comparison, which will give well-controlled results only if the true generative
model is in the list of models compared. \cite{friston2012} argues that,
based on the Neyman-Pearson lemma, cross-validation is less powerful than
likelihood ratio tests using the full dataset. However, it is
important to keep in mind that these approaches only test for
self-consistence, as the Neyman-Pearson lemma is established under the
hypothesis that the model used to define the test is indeed the 
data-generating process
\cite{neyman1933}, while in practice it is often the case that this model
gives poor fits to the data \cite{lohmann2012}. Applying test procedures
on different data than that used to fit the model, as in
cross-validation, is much more resilient to modeling errors.}.


\section{Conclusion}

Horwitz \emph{el al.}\ \cite{horwitz1995} claimed almost 20 years ago
that ``the crucial concept needed for network analysis is covariance''.
In our eyes, this still holds today. Estimation functional connectomes
relies largely on fitting covariance models. Their comparison requires
understanding how these covariances vary and finding metrics to capture
this variability. The additional secret ingredient may be using  
confounds regressors in all statistical steps. A good choice of a small number of
relevant regions facilitates connectome comparison. However, such a
choice cannot yet be fully factored out via methods and must rely on
neuroscientific expertise.

Methodological challenges to functional-connectome-based group studies
arise from the dimensionality and the variability of the connectome. With
the current tools, inter-subject comparison of connectomes comprising
many nodes is limited by the difficulty of estimating high-dimensional
covariance matrices and the loss of statistical power due to multiple
comparisons. Better algorithms integrating powerful a priori information are required
to push the limits of covariance estimation. Better characterization of
inter-subject variability of connectomes \cite{kelly2012} will help
choosing parameterizations and invariants to avoid testing each edge for
a difference, as this strategy inevitably leads to a needle in a haystack
problem.

Reviewing methodological options to learn and compare connectomes
highlights that there is currently no unique solution, but a spectrum of
related methods and analytical strategies. More empirical results are
required to guide the choices. However this diversity is probably
unavoidable: a diffuse disease like schizophrenia will not lead to the
same connectome modifications as a focal lesion. In statistical learning,
``no free lunch'' theorems \cite{wolpert1996} tell us that no strategy
can perform uniformly better in all situations. In practice, the key to a
successful analysis is to understand well the assumptions and
interpretation of each option, in order to match the method to the
question. Similarly, the idealized notion of an unique \emph{functional
connectome} to describe connections in brain function is probably an
utopia, and various connectomes should be considered in different
settings, such as the study of varying phenotypic conditions, or that of
on-going activity versus activity related to specific tasks.

\subsection*{Acknowledgments}

GV acknowledges funding from the NiConnect grant and the Dynamic
Diaschisis project DEQ20100318254 from \emph{Fondation pour la Recherche
M\'edicale}, as well as many insightful discussions with Andreas
Kleinschmidt on on-going activity and Bertrand Thirion on statistical
data processing. RCC would like to acknowledge support by a NARSAD Young
Investigator Grant from the Brain \& Behavior Research Foundation.
The authors would like to thank the anonymous reviewers
for their suggestions, which improved the manuscript.

{
\section*{References}
\small
\bibliographystyle{model1b-num-names}
\bibliography{biblio} }


\end{document}